\documentclass[copyright,creativecommons]{eptcs}
\providecommand{\event}{TARK 2017} 
\usepackage{breakurl}             
\usepackage{underscore}            
\usepackage[numbers]{natbib}

\usepackage[T1]{fontenc}
\usepackage[utf8]{inputenc}
\usepackage[ruled,vlined,noend]{algorithm2e}
\usepackage[noenvironments,noalghelp]{miritools}  
\usepackage{stmaryrd}                   
\SetSymbolFont{stmry}{bold}{U}{stmry}{m}{n}  
\usepackage{mleftright}            
\usepackage{dsfont}                
\usepackage{comment}               
\usepackage{float}                 
\usepackage{thmtools}              
\usepackage{thm-restate}           
\usepackage{bm}                    
\usepackage{booktabs}              
\usepackage[framemethod=tikz]{mdframed} 
\mdfsetup{skipabove=1em,skipbelow=1em}

\newtheorem{theorem}{Theorem}[subsection]
\newtheorem{definition}[theorem]{Definition}

\newenvironment{key}{\begin{mdframed}[innertopmargin=0.25em,innerbottommargin=0.75em]}{\end{mdframed}}
\newmdtheoremenv[innertopmargin=0.15em,nobreak=true]{keydef}[theorem]{Definition}

\numberwithin{equation}{subsection}
\newcounter{Hequation}

\makeatletter
\g@addto@macro\equation{\stepcounter{Hequation}}
\makeatother

\newtheorem{desideratum}{Desideratum}

\newenvironment{sketch}[1][]{\noindent\emph{Proof sketch.}\proofin{#1}\begin{quote}}{\end{quote}}

\newcommand{\Des}[1]{Desideratum~\ref{des:#1}}

\renewcommand{\iff}{\leftrightarrow}
\newcommand{\Implies}{\Rightarrow}

\newcommand{\fin}{\operatorname{Fin}}
\newcommand{\quot}[1]{``#1"}
\newcommand{\enc}[1]{{\underline{#1}}}
\newcommand{\seq}[1]{{\overline{#1}}}

\newcommand{\any}{-}

\newcommand{\prob}{p}

\newcommand{\price}{p}

\newcommand{\nn}{n}

\newcommand{\deff}{f}

\newcommand{\phis}{\seq{\phi}}
\newcommand{\psis}{\seq{\psi}}

\newcommand{\deltas}{\seq{\delta}}
\newcommand{\varepsilons}{\seq{{\varepsilon}}}

\newcommand{\probs}{\seq{{\prob}}}

\newcommand{\ec}[1][\ ]{e.c.#1}

\newcommand{\representscomputations}{can represent computable functions}

\newcommand{\li}{logical inductor}
\newcommand{\LItitle}{Logical Inductor}
\newcommand{\lic}{logical induction criterion}

\newcommand{\LICtitle}{Logical Induction Criterion}

\newcommand{\rparenthetical}[1]{\hfill\upshape{{\small (#1)}}} 
\newcommand{\proofin}[1]{}

\newcommand{\tinysectionend}[1]{\medskip\noindent}

\newcommand{\trade}{T}%

\newcommand{\Bayesian}{\mathrm{Pr}}
\newcommand{\BelState}{\mathbb{P}}

\newcommand{\Market}{\seq{\Pricing}}
\newcommand{\Pricing}{\mathbb{P}}

\newcommand{\pt}{\BelState}
\newcommand{\MP}{\seq{\pt}}

\newcommand{\dt}{D}
\newcommand{\DP}{\seq{\dt}}

\newcommand{\Theory}{\Gamma}
\newcommand{\Lang}{\mathcal{L}}
\newcommand{\Sentences}{\mathcal{S}}

\DeclareMathOperator{\OneOperator}{\mathds{1}} 
\newcommand{\ctsind}[1]{\operatorname{Ind}_{\text{\small{${#1}$}}}} 

\title{A Formal Approach to the Problem of Logical Non-Omniscience}

\author{Scott Garrabrant \qquad Tsvi Benson-Tilsen \qquad Andrew Critch  \\ Nate Soares \qquad Jessica Taylor
\institute{Machine Intelligence Research Institute \\ Berkeley, CA}
\email{\{scott,tsvi,critch,nate,jessica\}@intelligence.org}
}

\def\titlerunning{A Formal Approach to the Problem of Logical Non-Omniscience}
\def\authorrunning{S. Garrabrant, T. Benson-Tilsen, A. Critch, N. Soares, and J. Taylor}
\def\copyrightholders{Garrabrant, Benson-Tilsen, Critch, Soares \& Taylor}

\begin{document}
\maketitle

\begin{abstract}
 
We present the \emph{logical induction criterion} for computable algorithms that assign probabilities to every logical statement in a given formal language, and refine those probabilities over time. The criterion is motivated by a series of stock trading analogies. Roughly speaking, each logical sentence $\phi$ is associated with a stock that is worth \$1 per share if $\phi$ is true and nothing otherwise, and we interpret the belief-state of a logically uncertain reasoner as a set of market prices, where $\pt_\nn(\phi)=50\%$ means that on day $\nn$, shares of $\phi$ may be bought or sold from the reasoner for 50\textcent{}. A market is then called a \emph{logical inductor} if (very roughly) there is no polynomial-time computable trading strategy with finite risk tolerance that earns unbounded profits in that market over time. We then describe how this single criterion implies a number of desirable properties of bounded reasoners; for example, logical inductors outpace their underlying deductive process, perform universal empirical induction given enough time to think, and place strong trust in their own reasoning process.

\end{abstract}

\section{Introduction}\label{sec:intro}

Every student of mathematics has experienced uncertainty about conjectures for which there is ``quite a bit of evidence'', such as the Riemann hypothesis or the twin prime conjecture. Indeed, when Zhang \cite{zhang2014bounded} proved a bound on the gap between primes, we were tempted to increase our credence in the twin prime conjecture. But how much evidence does this bound provide for the twin prime conjecture? Can we quantify the degree to which it should increase our confidence?

The natural impulse is to appeal to probability theory in general and Bayes' theorem in particular. Bayes' theorem gives rules for how to use observations to update empirical uncertainty about unknown events in the physical world. 

However, probability theory lacks the tools to manage logical non-omniscience:
probability-theoretic reasoners cannot possess uncertainty about logical facts so long as their beliefs respect basic logical constraints. For example, let $\phi$ stand for the claim that the 87,653rd digit of $\pi$ is a 7. If this claim is true, then $(1+1=2) \Implies \phi$. But the laws of probability theory say that if $A \Implies B$ then $\Bayesian(A) \le \Bayesian(B)$. Thus, a perfect Bayesian must be at least as sure of $\phi$ as they are that $1+1=2$! Recognition of this problem dates at least back to \cite{Good:1950:weighing}.

Many have proposed methods for relaxing the criterion $\Bayesian(A) \le \Bayesian(B)$ until such a time as the implication has been proven (see, e.g., the work of \cite{Hacking:1967,Christiano:2014:omniscience}). But this leaves open the question of how probabilities should be assigned before the implication is proven, and this brings us back to the search for a principled method for managing uncertainty about logical facts when relationships between them are suspected but unproven.

In this paper we describe what we call the \emph{logical induction criterion} for reasoning under logical uncertainty.
Our solution works, more or less, by treating a reasoner's beliefs as prices in a market that fluctuate over time, and requiring that those prices not be exploitable indefinitely by any sequence of trades constructed by an efficient (polynomial-time) algorithm.
The logical induction criterion can be seen as a weakening of the ``no Dutch book'' criteria that Ramsey \cite{Ramsey:1931}, de Finetti \cite{DeFinetti:1937:foresight}, Teller \cite{teller1973conditionalization}, and Lewis \cite{lewis1999papers} used to support standard probability theory, which is analogous to the ``no Dutch book'' criteria that von Neumann and Morgenstern \cite{Von-Neumann:1944} and Joyce \cite{Joyce:1999} used to support expected utility theory. 
Because of the analogy, and the variety of desirable properties that follow immediately from this one criterion, we believe that the logical induction criterion captures a portion of what it means to do good reasoning about logical facts in the face of deductive limitations. 

\Sec{desiderata} lists desiderata for reasoning under logical uncertainty.

\Sec{relatedwork} lists further related work.

\Sec{framework} presents an overview of the logical induction framework.

\Sec{properties} discusses a collection of properties satisfied by logical inductors.

\Sec{discussion} gives concluding remarks.

Note on abridgement: Due to space considerations, this paper does not include proofs of claims, and describes some results only at a high level. The formal details of our definitions and theorems, additional properties of logical inductors, proofs of properties, a construction of a logical inductor, and further discussion can be found in \cite{Garrabrant:2016:li}.

\section{Desiderata for Reasoning under Logical Uncertainty}\label{sec:desiderata}
For historical context, and to further reify the problem, we now review a number of desiderata that have been proposed in the literature as desirable features of ``good reasoning'' in the face of logical uncertainty. 

\begin{desideratum}[Computable Approximability]\label{des:computable}\label{des:first}
  The method for assigning probabilities to logical claims (and refining them over time) should be computable.
\end{desideratum}

\begin{desideratum}[Coherence in the Limit]\label{des:coherent}\label{des:second} 
  The belief state that the reasoner is approximating better and better over time should be logically consistent.
\par\rparenthetical{Discussed in \Sec{limitprops}.}
\end{desideratum}

\begin{desideratum}[Approximate Coherence]\label{des:ic}
  The belief states of the reasoner over time should be approximately logically consistent. 
  \par\rparenthetical{Discussed in \Sec{timelylearning}.}
\end{desideratum}

\noindent \Des{ic} dates back to at least Good \cite{Good:1950:weighing}, who proposes a weakening of the condition of coherence that could apply to the belief states of limited reasoners. Hacking \cite{Hacking:1967} proposes an alternative weakening, as do Garrabrant et al. \cite{Garrabrant:2016:ic}.

\begin{desideratum}[Learning of Statistical Patterns]\label{des:stats}
  In lieu of knowledge that bears on a logical fact, a good reasoner should assign probabilities to that fact in accordance with the rate at which similar claims are true.
\end{desideratum}

\noindent For example, a good reasoner should assign probability $\approx 10\%$ to the claim ``the $n$th digit of $\pi$ is a 7'' for large $n$ (assuming there is no efficient way for a reasoner to guess the digits of $\pi$ for large $n$); see \cite{Savage:1967:personal}.

\begin{desideratum}[Calibration]\label{des:calibration}
  Good reasoners should be well-calibrated. That is, among events that a reasoner says should occur with probability $p$, they should in fact occur about $p$ proportion of the time.
\end{desideratum}

\begin{desideratum}[Non-Dogmatism]\label{des:nondogmatism}
  A good reasoner should not have extreme beliefs about mathematical facts, unless those beliefs have a basis in proof.
\par\rparenthetical{Discussed in \Sec{limitprops}.}
\end{desideratum}

\noindent In the domain of logical uncertainty, \Des{nondogmatism} can be traced back to Carnap \cite[Sec. 53]{Carnap:1962:LogicalProbability}, and has been demanded by many, including Gaifman\cite{Gaifman:1982:RichProbabilities} and Hutter \cite{Hutter:2013}.

\begin{desideratum}[Uniform Non-Dogmatism]\label{des:pa}
  A good reasoner should assign a non-zero probability to any computably enumerable consistent theory (viewed as a limit of finite conjunctions).
  \par\rparenthetical{Discussed in \Sec{limitprops}.}
\end{desideratum}

\noindent  The first formal statement of \Des{pa} that we know of is given by Demski \cite{Demski:2012a}, though it is implicitly assumed whenever asking for a set of beliefs that can reason accurately about arbitrary arithmetical claims (as is done by, e.g., Savage \cite{Savage:1967:personal} and Hacking \cite{Hacking:1967}).

\begin{desideratum}[Universal Inductivity]\label{des:solomonoff}
  Given enough time to think, the beliefs of a good reasoner should dominate any (lower semicomputable) semimeasure.
  \par\rparenthetical{Discussed in \Sec{limitprops}.}
\end{desideratum}

\begin{desideratum}[Approximate Bayesianism]\label{des:bayes}
  The reasoner's beliefs should admit of some notion of conditional probabilities, which approximately satisfy both Bayes' theorem and the other desiderata listed here.
\end{desideratum}

\begin{desideratum}[Self-knowledge]\label{des:introspection}
  If a good reasoner knows something, she should also know that she knows it.
  \rparenthetical{Discussed in \Sec{introspection}.}
\end{desideratum}

\noindent Proposed by Hintikka \cite{Hintikka:1962:knowledge}, \Des{introspection} is popular among epistemic logicians. This desideratum has been formalized in many different ways; see \cite{Christiano:2013:definability,Campbell:2015:SelfReference} for a sample.

\begin{desideratum}[Self-Trust]\label{des:lob}\label{des:penult}
  A good reasoner thinking about a hard problem should expect that, in the future, her beliefs about the problem will be more accurate than her current beliefs.
  \rparenthetical{Discussed in \Sec{selftrust}.}
\end{desideratum}

\begin{desideratum}[Approximate Inexploitability]\label{des:inexp}\label{des:last}
  It should not be possible to run a Dutch book against a good reasoner in practice.
  \rparenthetical{See \Sec{criterion} for our proposal.}
\end{desideratum}
  
\noindent As noted by Eells \cite{Eells:1990:OldEvidence}, the Dutch book constraints used by von Neumann and Morgenstern \cite{Von-Neumann:1944} and de Finetti \cite{DeFinetti:1937:foresight} are implausibly strong: all it takes to run a Dutch book according to de Finetti's formulation is for the bookie to know a logical fact that the reasoner does not know. Thus, to avoid being Dutch booked by de Finetti's formulation, a reasoner must be logically omniscient.

Hacking \cite{Hacking:1967} and Eells \cite{Eells:1990:OldEvidence} call for weakenings of the Dutch book constraints, in the hopes that reasoners that are approximately inexploitable would do good approximate reasoning. This idea is the cornerstone of our framework---we consider reasoners that cannot be exploited by betting strategies that can be constructed by a polynomial-time machine.

Logical inductors 
satisfy desiderata~\ref{des:first} through~\ref{des:last}. In fact, logical inductors are designed to meet only~\Des{computable} (computable approximability) and~\Des{inexp} (approximate inexploitability), from which~\ref{des:second}-\ref{des:penult}  all follow (see \cite{Garrabrant:2016:li}).

\section{Additional Related Work}\label{sec:relatedwork}

The study of logical uncertainty is an old topic. It can be traced all the way back to Bernoulli, who laid the foundations of statistics, and later Boole \cite{boole1854investigation}, who was interested in the unification of logic with probability from the start. Refer to \cite{hailperin1996sentential} for a historical account. Our algorithm assigns probabilities to sentences of logic directly; this thread can be traced back through {\L}o{\'{s} \cite{Los:1955} and later Gaifman \cite{Gaifman:1964}, who developed the notion of coherence that we use in this paper.

When it comes to the problem of developing formal tools for manipulating uncertainty, our methods are heavily inspired by Bayesian probability theory, and so can be traced back to Pascal, who was followed by Bayes, Laplace, Kolmogorov \cite{kolmogorov1950foundations}, Savage \cite{savage1954foundations}, Carnap \cite{Carnap:1962:LogicalProbability}, Jaynes \cite{Jaynes:2003}, and many others. Polya \cite{polya1990mathematics} was among the first in the literature to explicitly study the way that mathematicians engage in plausible reasoning, which is tightly related to the object of our study.

In addition to Good \cite{Good:1950:weighing}, Savage \cite{Savage:1967:personal}, and Hacking \cite{Hacking:1967}, the flaw in Bayesian probability theory was also highlighted by Glymour \cite{Glymour:1980:OldEvidence}, and dubbed the ``problem of old evidence'' by Garber \cite{Garber:1983:OldEvidence} in response to Glymor's criticism. Eells \cite{Eells:1990:OldEvidence} gave a lucid discussion of the problem, revealed flaws in Garber's arguments and in Hacking's solution, and named a number of other desiderata which our algorithm manages to satisfy; see \cite{zynda1995old} and \cite{sprenger2015novel}. Adams \cite{adams1996primer} uses logical deduction to reason about an unknown probability distribution that satisfies certain logical axioms. Our approach works in precisely the opposite direction: we use probabilistic methods to create an approximate distribution where logical facts are the subject.

Some work in epistemic logic has been directed at modeling the dynamics of belief updating in non-omniscient agents; see for example \cite{konolige1983deductive,velazquez2014dynamic,balbiani2016logical}. Our approach differs in that we use first-order logic, and therefore use the recursion theorem to make reflective statements instead of using explicit knowledge or belief operators; the potential paradoxes of self-reference are circumvented by allowing beliefs to be probabilistic. The mechanism used by our logical inductor to update its beliefs is not very transparent, leaving open the possibility of a more principled understanding of the local mechanics of updating probabilities on logical or inductive inferences.

Straddling the boundary between philosophy and computer science, Aaronson \cite{Aaronson:2013:PhilosophersComplexity} has made a compelling case that computational complexity must play a role in answering questions about logical uncertainty. Fagin and Halpern \cite{fagin1987belief} also straddled this boundary with early discussions of algorithms that manage uncertainty in the face of resource limitations. (See also their discussions of uncertainty and knowledge. \cite{Fagin:1995:knowledge,Halpern:2003})

\section{The \LICtitle}\label{sec:framework}\label{sec:criterion}

We propose a partial solution to the problem of logical non-omniscience, which we call \emph{logical induction}. Roughly speaking, a \emph{logical inductor} is a computable reasoning process that is not exploitable by any polynomial-time computable strategy for making trades against it, using its probabilities as the prices of shares. In this section we give a high-level overview  of the criterion and the main result (details are in \cite{Garrabrant:2016:li}), before giving precise statements in \Sec{properties} of some of the properties satisfied by logical inductors.

Very roughly, our setup works as follows. We consider reasoners that assign probabilities to sentences $\Sentences$ written in some formal language  $\Lang$.

\begin{definition}[Pricing]\label{def:pricing}
  A \textbf{pricing} is a  computable rational function $\Pricing : \Sentences \to \QQ \cap [0, 1]$.
\end{definition}

\noindent Here $\Pricing(\phi)$ is interpreted as the probability of~$\phi$.  We can visualize  a pricing as a list of $(\phi, \prob)$ pairs, where the $\phi$ are unique sentences and the $\prob$ are rational-number probabilities, and $\Pricing(\phi)$ is defined to be $\prob$ if $(\phi, \prob)$ occurs in the list, and $0$ otherwise. (In this way we can represent belief states of reasoners that can be written down explicitly in a finite amount of space.) The output of a reasoner is then nothing but a sequence of pricings:
 
\begin{definition}[Market]\label{def:marketprocess}
  A \textbf{market} $\Market=(\Pricing_1,\Pricing_2,\ldots)$ is a computable sequence of pricings $\Pricing_i : \Sentences \to \QQ \cap [0,1]$.
\end{definition}

\noindent The pricings $(\Pricing_1,\Pricing_2,\ldots)$ represent the belief states of a reasoner progressively refining their opinions about the logical statements in $\Sentences$.  In the background, there is some process producing progressively larger sets of trusted statements:

\begin{definition}[Deductive Process]\label{def:dedproc}
  A \textbf{deductive process} $\DP : \NN^+ \to \fin(\Sentences)$ is a computable nested sequence $\dt_1 \subseteq \dt_2 \subseteq \dt_3 \ldots$ of finite sets of sentences.  
\end{definition}

\noindent The deductive process $\DP$ can be thought of as a theorem prover for some trusted logical theory $\Theory$ in the language $\Lang$. Indeed, we will henceforth assume that $\Theory = \bigcup_\nn \dt_\nn$. 
Thus the goal of our reasoner $\Market$ is to anticipate which statements will be proven or disproven by  $\Theory$, well before the rote proof-search $\DP$ decides those statements.

As in classical Dutch book arguments for probability theory, in addition to seeing $\Pricing(\phi)=\price$ as an assignment of subjective credence to $\phi$, we also view $\Pricing(\phi)$ as a stance with respect to which bets are desirable or not. That is, we interpret $\Pricing(\phi)=\price$ to mean that the price of a $\phi$-share according to $\Pricing$ is $\price$, where (roughly speaking) a $\phi$-share is worth \$1 if $\phi$ is true. This allows us to set up Dutch book arguments against a reasoner using computable bookies:

\begin{definition}[Trader]\label{def:trader}
  A \textbf{trader} is a sequence $(\trade_1, \trade_2, \ldots)$ where each $\trade_\nn$ is a trading strategy for day $\nn$.
\end{definition}

\noindent Without belaboring the details, a trading strategy for day $n$ is a strategy for responding to the day's market prices $\Pricing_n$ with buy orders and sell orders for  shares in sentences from $\Sentences$. (Formally, it is a continuous function from pricings to linear combinations of sentences, expressed in some computable language.) Over time, a trader accumulates cash and stock holdings from the trades it makes against $\MP$.

The logical induction criterion then demands of market prices $\MP$ that no efficiently computable trader can reliably make money by trading against  the market prices $(\Pricing_1,\Pricing_2,\ldots)$:

\begin{key}
\begin{restatable}[The Logical Induction Criterion]{definition}{criterion}\label{def:lic}
  A market $\MP$ is said to satisfy the \textbf{\lic{}} relative to a deductive process $\DP$ if there is no efficiently computable trader  that exploits $\MP$ relative to $\DP$.  A market $\MP$ meeting this criterion is called a \textbf{\li{} over $\bm{\DP}$}.
\end{restatable}
\end{key}

\noindent Again glossing over details, a trader is said to exploit $\MP$ relative to $\DP$ if the possible values of the trader's holdings from trading against $\MP$ are unboundedly high over time, without being unboundedly low, where holdings are evaluated by what truth assignments to $\Sentences$ are propositionally consistent with $\dt_n$ at time $n$. 
Here, ``efficiently computable'' (abbreviated e.c.) can be taken to mean computable in time polynomial in $n$, but this is not crucial to the definition. Given the assumption that $\Theory = \bigcup_\nn \dt_\nn$, we also say that $\MP$ is a logical inductor over $\Theory$. 

Our key theorem is that this criterion, while gratifyingly  strong, is also feasible:

\begin{key}
\begin{restatable}{theorem}{logindcri}\label{thm:li}
  For any deductive process $\DP$, there exists a computable belief sequence $\MP$ satisfying the \lic{} relative to $\DP$.
\end{restatable}
\end{key}

\section{Properties of \LItitle{}s}\label{sec:properties}

Here is an intuitive argument that logical inductors perform good reasoning under logical uncertainty:

\begin{quote}
  Consider any polynomial-time method for efficiently identifying patterns in logic. If the market prices don't learn to reflect that pattern, a clever trader can use that pattern to exploit the market. Thus, a logical inductor must learn to identify those patterns.
\end{quote}

\noindent This section will substantiate this argument by stating a number of properties satisfied by logical inductors, corresponding to some of the desiderata discussed in \Sec{desiderata}. Proofs of \Thm{li} and the theorems in this section can be found in \cite{Garrabrant:2016:li}.

\subsection{Notation}
Throughout, we assume that $\MP$ is a logical inductor over the theory $\Theory$. We also assume that $\Theory$ represents computations in the technical sense, i.e. we can write terms in $\Lang$ that stand for computations, and $\Theory$ proves that those terms evaluate to their correct value (and no other value). 

We will enclose sentences in quotation marks when they are used as syntactic objects. An underlined symbol should be replaced by the expression it stands for. For example, $\enc{\deff}(\enc{\nn})$ stands for a program that computes the function $f$ given input $n$, whereas $\enc{\deff(\nn)}$ stands for the numeral $f(n)$ evaluates to.

We use an  overline to denote sequences of sentences, probabilities, and other objects, as in $\MP$ and $\DP$; for example, $\phis$ is the sequence of sentences $(\phi_1,\phi_2, \dots)$.  A sequence $\seq{x}$ is efficiently computable (e.c.) if and only if there exists a computable function $\nn \mapsto x_\nn$ with runtime polynomial in $\nn$.
Given any sequences $\seq x$ and $\seq y$, we write 
\begin{align*}
  x_\nn \eqsim_\nn y_\nn & \quad\text{for}\quad \lim_{n \to\infty} x_\nn - y_\nn = 0,\text{~and}\\
  x_\nn \gtrsim_\nn y_\nn & \quad\text{for}\quad \liminf_{n \to\infty} x_\nn - y_\nn \ge 0.
\end{align*}

\subsection{Properties of the limit}\label{sec:limitprops}

Firstly, the market prices of a \li{} converge:

\begin{restatable}[Convergence]{theorem}{convergence}\label{thm:con}\label{thm:first}
  The limit ${\pt_\infty:\Sentences\rightarrow[0,1]}$ defined by \[\pt_\infty(\phi) := \lim_{\nn\rightarrow\infty} \pt_\nn(\phi)\] exists for all $\phi$.
\end{restatable}

\begin{restatable}{sketch}{sketchcon}
  Roughly speaking, if $\MP$ never makes up its mind about $\phi$, then it can be exploited by a trader arbitraging shares of $\phi$ across different days. That is,  suppose by way of contradiction that  $\pt_\nn(\phi)$ never settles down, but rather oscillates by a substantial amount infinitely often. Then there is a trader that repeatedly buys a share in $\phi$ when the price is low, and sells it back when the price is high. This trader accumulates unbounded wealth, thereby exploiting $\MP$, which contradicts that $\MP$ is a logical inductor; therefore the limit $\pt_\infty(\phi)$ must in fact exist.
\end{restatable}

\noindent This sketch showcases the main intuition for the convergence of $\MP$, but elides a number of crucial details; see \cite{Garrabrant:2016:li}.

Next, the limiting beliefs of a \li{} represent a coherent probability distribution:

\begin{restatable}[Limit Coherence]{theorem}{limitcoherence}\label{thm:lc}
  $\pt_\infty$ is coherent, i.e., it gives rise to an internally consistent probability measure $\Bayesian$ on the set of all consistent completions $\Theory':\Sentences\to\BB$ of~$\Theory$, defined by the formula \[\Bayesian(\Theory'(\phi)=1):=\pt_\infty(\phi).\]
\end{restatable}

\noindent First formalized by Gaifman \cite{Gaifman:1964}, coherence says that beliefs should satisfy probabilistic versions of logical consistency; for example, the reasoner should assign  
 $\Bayesian(\phi) \leq \Bayesian(\psi)$ if $\phi \Implies \psi$, etc.
 This theorem is proven using methods analogous to standard Dutch book arguments for coherent beliefs, translated into the language of traders.


Convergence and coherence together justify that a logical inductor $\MP$ approximates a belief state that is consistent with the background theory $\Theory$. What else is there to say about the limiting beliefs $\pt_\infty$ of a logical inductor?

For starters,  $\MP$ learns not to assign extreme probabilities to sentences that are independent from $\Theory$:

\begin{restatable}[Non-Dogmatism]{theorem}{restatenondog}\label{thm:nd}
 If $\Theory \nvdash \phi$ then 
  $\pt_\infty(\phi)<1$, and if $\Theory \nvdash \neg\phi$ then $\pt_\infty(\phi)>0$.
\end{restatable}

\noindent Non-dogmatism can be viewed as an inductive property: non-dogmatic beliefs can be easily conditioned on events (sentences) that haven't already been observed (proved or disproved), producing a coherent conditional belief state, whereas conditioning dogmatic beliefs can cause problems. 

We can push the idea of inductive reasoning much further, following the work of  Solomonoff \cite{Solomonoff:1964,Solomonoff:1964a}, Zvonik and Levin \cite{zvonkin1970complexity} and Li and Vit\'anyi \cite{Li:1993} on empirical sequence prediction. They describe an inductive process (known as a universal semimeasure) that predicts as well or better than any computable predictor, modulo a constant amount of error.
Although universal semimeasures are uncomputable, we can ask logically uncertain reasoners to copy those successes given enough time to think:

\begin{restatable}[Domination of the Universal Semimeasure]{theorem}{restatedus}\label{thm:dus}
  Let $(b_1, b_2, \ldots)$ be a sequence of zero-arity predicate symbols in $\Lang$ not mentioned in $\Theory$, and let $\sigma_{\le n}=(\sigma_1,\ldots,\sigma_\nn)$ be any finite bitstring. Define
  \[
    \pt_\infty(\sigma_{\le n}) :=
    \pt_\infty(\quot{(b_1 \iff \enc{\sigma_1}=1) \land
    \ldots \land (b_n \iff \enc{\sigma_n}=1)}),
  \]
  such that, for example, $\pt_\infty(01101) = \pt_\infty(\quot{\lnot b_1 \land b_2 \land b_3 \land \lnot b_4 \land b_5})$.
  Let $M$ be a universal continuous semimeasure. Then there is some positive constant $C$ such that for any finite bitstring $\sigma_{\le n}$,
  \[
    \pt_\infty(\sigma_{\le n}) \ge C \cdot M(\sigma_{\le n}).
  \]
  \proofin{\ref{app:dus}}
\end{restatable}

\noindent 
In other words, logical inductors are a computable approximation to a normalized probability distribution that dominates any lower semicomputable semimeasure. In fact, this dominance is strict: $\pt_\infty$ will e.g., assign positive probability to sequences that encode completions of Peano arithmetic, which the universal semimeasure does not do.\footnote{This does not contradict the universality of $M$, as $\pt_\infty$ is higher in the arithmetical hierarchy than $M$.}

\subsection{Outpacing deduction}\label{sec:timelylearning}

It is not too difficult to define a reasoner that assigns probability~1 to all (and only) the provable sentences, in the limit: simply assign probability 0 to all sentences, and then enumerate all logical proofs, and assign probability~1 to the proven sentences. The real trick is to recognize patterns in a timely manner, well before the sentences can be proven by slow deduction.

\begin{restatable}[Provability Induction]{theorem}{restatepi}\label{thm:provind}\label{thm:patfirst}
  Let $\phis$ be an \ec sequence of theorems. Then
  \[
    \pt_\nn(\phi_\nn) \eqsim_\nn 1.
  \]
  Furthermore, let $\psis$ be an \ec sequence of disprovable sentences. Then
  \[
    \pt_\nn(\psi_\nn) \eqsim_\nn 0.
  \]
\end{restatable}
\begin{sketch}[\ref{sec:provind} or~\ref{app:provind}]
Suppose not. Then there is a trader that buys a share in $\phi_n$ whenever the price is too far below \$1, and then waits for $\phi_n$ to appear in the deductive process $\DP$, repeating this process indefinitely. This trader would exploit $\MP$, a contradiction.

\end{sketch}

\noindent In other words, $\MP$ will learn to start believing $\phi_\nn$ by day $\nn$ at the latest, despite the fact that $\phi_\nn$ won't be deductively confirmed until day $f(\nn)$, which is potentially much later. In colloquial terms, if $\phis$ is a sequence of facts that can be generated efficiently, then $\MP$ inductively learns the pattern, and its belief in $\phis$ becomes accurate faster than $\DP$ can computationally verify the individual sentences.

\begin{quote}
  \textbf{Analogy: Ramanujan and Hardy.} Imagine that the statements $\phis$ are being output by an algorithm that uses heuristics to generate mathematical facts without proofs, playing a role similar to the famously brilliant, often-unrigorous mathematician Srinivasa Ramanujan. Then $\MP$ plays the historical role of the beliefs of the rigorous G.H.\ Hardy who tries to verify those results according to a slow deductive process ($\smash{\DP}$). After Hardy ($\MP$) verifies enough of Ramanujan's claims ($\phi_{\le \nn}$), he begins to trust Ramanujan, even if the proofs of Ramanujan's later conjectures are incredibly long, putting them ever-further beyond Hardy's current abilities to rigorously verify them. In this story, Hardy's inductive reasoning (and Ramanujan's also) outpaces his deductive reasoning.
\end{quote}

\noindent To further emphasize the meaning of \Theorem{provind}, consider the famous halting problem of Turning \cite{turing1936computable}. Turing proved that there is no general algorithm for determining whether or not an arbitrary computation halts. Let's examine what happens when we confront logical inductors with the halting problem.

\begin{restatable}[Learning of Halting Patterns]{theorem}{restatehalts}\label{thm:halts}
  Let $\seq{m}$ be an \ec sequence of Turing machines, and $\seq{x}$ be an \ec sequence of bitstrings, such that $m_\nn$ halts on input $x_\nn$ for all $\nn$. Then
  \[
    \pt_\nn(\quot{\text{$\enc{m_\nn}$ halts on input $\enc{x_\nn}$}}) \eqsim_\nn 1.
  \]
  \proofin{\ref{app:halts}}
\end{restatable}

\noindent Of course, this is not so hard on its own---a function that assigns probability~1 to everything also satisfies this property. The real trick is separating the halting machines from the non-halting ones. 

By undecidability, there are  Turing machines~$q$ that fail to halt on input~$y$, but such that $\Theory$ is not strong enough to prove this fact. In this case, $\pt_\infty$'s probability of~$q$ halting on input~$y$ is positive, by \Theorem{nd}. Nevertheless, $\MP$ still learns to stop expecting that those machines will halt after any reasonable amount of time:

\begin{restatable}[Learning not to Anticipate Halting]{theorem}{restatedontwait}\label{thm:dontwait}
  Let $\seq{q}$ be an \ec sequence of Turing machines, and let $\seq{y}$ be an \ec sequence of bitstrings, such that $q_\nn$ does not halt on input $y_\nn$ for any $\nn$. Let $f$ be any computable function. Then
  \[
    \pt_\nn(\quot{\text{$\enc{q_\nn}$ halts on input $\enc{y_\nn}$ within $\enc{f}(\enc{\nn})$ steps}}) \eqsim_\nn 0.
  \]
  \proofin{\ref{app:dontwait}}
\end{restatable}

\noindent These theorems can be interpreted as justifying the intuitions that many computer scientists have long held towards the halting problem: It is impossible to tell whether or not a Turing machine halts in full generality, but for large classes of well-behaved computer programs (such as \ec sequences of halting programs and provably non-halting programs) it's quite possible to develop reasonable and accurate beliefs. The boundary between machines that compute fast-growing functions and machines that never halt is difficult to distinguish, but even in those cases, it's easy to learn to stop expecting those machines to halt within any reasonable amount of time. 

As a consequence of of \Theorem{dontwait}, a logical inductor will trust their (computable) underlying deductive process $\DP$ to remain consistent for arbitrarily long specified  periods of time, if in fact $\DP$ is consistent. In other words, a logical inductor over the theory $\Theory$ will learn trust in the finitary  consistency of $\Theory$.

One possible objection here is that the crux of the halting problem (and of the $\Theory$-trust problem) is not about making good predictions, it is about handling diagonalization and paradoxes of self-reference.
So let us turn to the topic of $\MP$'s beliefs about $\MP$ itself.

\subsection{Self-knowledge}\label{sec:introspection}

Because we're assuming $\Theory$ \representscomputations, we can write sentences describing the beliefs of $\MP$ at different times. What happens when we ask $\MP$ about sentences that refer to itself?

\begin{restatable}[Self-knowledge]{theorem}{restateref}\label{thm:ref}
  Let $\phis$ be an \ec sequence of sentences, let $\seq{a}$, $\seq{b}$ be \ec sequences of probabilities. Then, for any \ec sequence of positive rationals $\deltas \to 0$, there exists a sequence of positive rationals $\varepsilons \to 0$ such that for all $\nn$:
  \begin{enumerate}    
    \item if $\pt_\nn(\phi_\nn)\in(a_\nn+\delta_\nn,b_\nn-\delta_\nn)$, then
    \[
      \pt_\nn(\quot{\enc{a_\nn} < \enc{\pt}_\enc{\nn}(\enc{\phi_\nn}) < \enc{b_\nn}}) > 1-\varepsilon_\nn,
    \] 
    \item if $\pt_\nn(\phi_\nn)\notin(a_\nn-\delta_\nn,b_\nn+\delta_\nn)$, then
    \[
      \pt_\nn(\quot{\enc{a_\nn} < \enc{\pt}_\enc{\nn}(\enc{\phi_\nn}) < \enc{b_\nn}}) < \varepsilon_\nn.
    \] 
  \end{enumerate}
  \proofin{\ref{app:ref}}
\end{restatable}

\noindent In other words, for any pattern in $\MP$'s beliefs that can be efficiently written down (such as ``$\MP$'s probabilities on $\phis$ are between $a$ and $b$ on these days''), $\MP$ learns to believe the pattern if it's true, and to disbelieve it if it's false (with vanishing error).  (Recall that the underlines indicate that the underlined expression should be expanded to the appropriate logical formula or term, representing e.g., the source code of an algorithm implementing $\MP$.)

At a first glance, this sort of self-reflection may seem to make \li{}s vulnerable to paradox. For example, consider the sequence of sentences $\seq{\chi^{0.5}}$ defined using the diagonal lemma by
\[
 \chi^{0.5}_n := \quot{{\enc{\pt}_{\enc{\nn}}}(\enc{\chi^{0.5}_\nn}) < 0.5} 
\]
such that $\chi^{0.5}_\nn$ is true iff $\MP$ assigns it a probability less than 50\% on day $\nn$. Such a sequence can be defined by G\"odel's diagonal lemma. These sentences are probabilistic versions of the classic ``liar sentence'', which has caused quite a ruckus in the setting of formal logic \cite{grim1991incomplete,mcgee1990truth,glanzberg2001liar,gupta1993revision,eklund2002inconsistent}. Because our setting is probabilistic, it's perhaps most closely related to the ``unexpected hanging'' paradox---$\chi^{0.5}_\nn$ is true iff $\MP$ thinks it is unlikely on day $\nn$. How do logical inductors handle this sort of paradox?

\begin{restatable}[Paradox Resistance]{theorem}{restatelp}\label{thm:lp}
  Fix a rational $\prob\in(0,1)$, and define an \ec sequence of ``paradoxical sentences'' $\seq{\chi^\prob}$ satisfying
  \[
    \Theory \vdash{{\enc{\chi^\prob_\nn}} \iff \left(
      {\enc{\pt}_{\enc{\nn}}}({\enc{\chi^\prob_\nn}}) < \enc{\prob}
    \right)}
  \]
  for all $\nn$. Then
  \[
    \lim_{\nn\to\infty}\pt_\nn(\chi^\prob_\nn)=\prob.
  \]
  \proofin{\ref{app:lp}}
\end{restatable}

\noindent In words, a logical inductor responds to paradoxical sentences $\seq{\chi^\prob}$ by assigning them probabilities that converge on $\prob$.

To understand why this is desirable, imagine that your friend owns a high-precision brain-scanner and can read off your beliefs. Imagine they ask you what probability you assign to the claim ``you will assign probability $<$80\% to this claim at precisely 10am tomorrow''. As 10am approaches, what happens to your belief in this claim? If you become extremely confident that it's going to be true, then your confidence should drop. But if you become fairly confident it's going to be false, then your confidence should spike. Thus, your probabilities should oscillate, pushing your belief so close to 80\% that you're not quite sure which way the brain scanner will actually call the claim, and you think the scanner is roughly 80\% likely to call it true. In response to a paradoxical claim, this is exactly how $\MP$ behaves, once it's learned how the paradoxical sentences work.

\subsection{Self-Trust}\label{sec:selftrust}
We've seen that logical inductors can, without paradox, have accurate beliefs about their own current beliefs.
Next, we turn our attention to the question of what a \li{} believes about its \emph{future} beliefs.

The coherence conditions of classical probability theory guarantee that, though a probabilistic reasoner expects their future beliefs to change in response to new empirical observations, they don't e.g., believe that their  future credence in $\phi$ is, in net expectation, lower than their current credence in $\phi$. 
For example, if a reasoner $\Bayesian(\any)$ knows that tomorrow they'll see some evidence $e$ that will convince them that Miss Scarlet was the murderer, then they already believe that she was the murderer today:
\[
  \Bayesian(\mathrm{Scarlet}) = \Bayesian(\mathrm{Scarlet}\mid e) \Bayesian(e) + \Bayesian(\mathrm{Scarlet}\mid \lnot e) \Bayesian(\lnot e).
\]
In colloquial terms, this says ``my current beliefs are \emph{already} a mixture of my expected future beliefs, weighted by the probability of the evidence that I expect to see.''

Logical inductors obey similar coherence conditions with respect to their future beliefs, with the difference being that a logical inductor updates its belief by gaining more knowledge about \emph{logical} facts, both by observing an ongoing process of deduction and by thinking for longer periods of time.  

To refer to $\MP$'s \emph{expectations} about its future self, we need a notion of logically uncertain variables. To avoid needless detail, suffice it to say that logically determined quantities, such as the output of a given computer program, can be represented and manipulated analogously to random variables in probability theory. We can write these variables as terms representing their value; for example, the variable written $\quot{\enc{\pt}_{\enc{\nn}}(\enc{\phi})}$  represents the probability assigned to $\phi$ by $\MP$ on day $n$. Using the beliefs $\pt_n$ of $\MP$ about $X$ on day $n$, we can define the (approximate) expectation $\EE_\nn(X)$.

We also need to know which future self our logical inductor will defer to:

\begin{definition}[Deferral Function]\label{def:deferralfunc}
  A function $\deff : \NN^+ \to \NN^+$ is called a \textbf{deferral function} if
  \begin{enumerate}
    \item $\deff(\nn) > \nn$ for all $\nn$, and
    \item as a function of $n$, $\deff(\nn)$ can be computed in time polynomial in $\deff(\nn)$.
  \end{enumerate}
\end{definition}

\noindent Now we can state the sense in which logical inductors don't expect, on net,  their future beliefs to be wrong in any particular direction.
\begin{restatable}[No Expected Net Update]{theorem}{restateceu}\label{thm:ceu}
  Let $\deff$ be a deferral function, and let $\phis$ be an \ec sequence of sentences. Then
  \[
      \pt_\nn(\phi_\nn) \eqsim_\nn 
      \EE_\nn(\quot{\enc{\pt}_{\enc{\deff}(\enc{\nn})}(\enc{\phi_\nn})}).
  \]
  \proofin{\ref{app:ceu}}
\end{restatable}

\noindent This theorem only says  that $\pt_n$ doesn't expect the beliefs of $\pt_{f(n)}$ about $\phis$ to err in a particular direction. A priori, it is possible that $\pt_n$ nevertheless believes its future beliefs $\pt_{f(n)}$ will be grossly misguided. For example, suppose that $\pt_n$ is very confident that $\pt_{f(n)}$ will have sufficient time to compute the truth of $\phi$, but will react insanely to this information: 
\[\pt_n(\quot{\enc{\pt}_{\enc{\deff}(\enc{\nn})}(\enc{\phi})=0} \mid \phi) = 1 
\]
\noindent and
\[\pt_n(\quot{\enc{\pt}_{\enc{\deff}(\enc{\nn})}(\enc{\phi})=1} \mid \lnot\phi) = 1.
\]
This is a priori consistent with \Thm{ceu} so long as $\pt_n$ assigns $\pt_n(\phi) = 0.5$, but it clearly indicates that $\pt_n$ does not trust its future beliefs.

To instead formalize the idea of a reasoner $\Bayesian$ that trusts their own reasoning process, let us first consider a self-trust property  in the setting of deductive logic:
\[
    \vdash \square \phi \to \phi.
\]
This property of deductive systems says that the system proves ``If I prove $\phi$ at some point, then it is true''. However, any sufficiently strong reasoner that satisfies this property for the statement $\phi=\bot$ is inconsistent by G\"{o}del's second incompleteness theorem! 
The search for logics that place confidence in their own machinery
dates at least back to Hilbert \cite{Hilbert:1902}. While G\"{o}del et al. \cite{Godel:1934} dashed these hopes, it is still desirable for reasoners to trust their reasoning process relatively well, most of the time (which humans seem to do).

As discussed in \Sec{timelylearning}, logical inductors trust their underlying deductive process $\DP$ in a slightly weaker, finitary sense. More interestingly, it turns out that logical inductors also trust their own reasoning process as a whole, including their inductive conclusions, in a manner that we now formalize.

Instead of $\vdash \square \phi \to \phi$, we can replace provability with high confidence, and then ask for something like
\begin{equation*} \label{eq:st}
  \Bayesian_\mathrm{now}(\phi \mid \Bayesian_\mathrm{later}(\phi) > \prob) \gtrsim \prob.
\end{equation*}
Colloquially, this says that if we tell $\Bayesian$ that in the future they will place more than $\prob$ credence in $\phi$, then they update their current beliefs to place at least $\prob$ credence. In short, $\Bayesian$ trusts that the outputs of their own ongoing reasoning process will be accurate.

Now, in fact property~\ref{eq:st} is not quite desirable as stated (and logical inductors do not satisfy it). Indeed, consider the liar sentence $\chi^\prob$ defined by
\[
 \chi^{\prob} := \quot{  \Bayesian_{\mathrm{later}} (\chi^{\prob}) < \prob}.
\]
A good reasoner will then satisfy 
\[
  \Bayesian_\mathrm{now}(\chi^{\prob} \mid \Bayesian_\mathrm{later}(\chi^{\prob}) > \prob) \eqsim 0,
\]
contradicting equation~\ref{eq:st}. The issue is that if we give $\Bayesian_\mathrm{now}$ high-precision access to the probabilities assigned by $\Bayesian_\mathrm{later}$---for example by conditioning on them---then  $\Bayesian_\mathrm{now}$ can outperform the (unconditioned) beliefs of $\Bayesian_\mathrm{later}$, in this case by having correct opinions about the liar sentence for $\Bayesian_\mathrm{later}$. 

Instead, we have the following self-trust property, which only gives $\pt_n$ limited-precision access to the beliefs of $\pt_{f(n)}$:

\begin{restatable}[Self-Trust]{theorem}{restatest}\label{thm:st}
  Let $\deff$ be a deferral function, $\phis$ be an \ec sequence of sentences, $\deltas$ be an \ec sequence of positive rational numbers, and $\probs$ be an \ec sequence of rational probabilities. Then
\begin{multline*}
    \EE_\nn\left(\quot{
      \enc{\OneOperator(\phi_\nn)} \cdot
      \enc{\ctsind{\delta_\nn}}\mleft(
        \enc{\pt}_{\enc{\deff}(\enc{\nn})}(\enc{\phi_\nn}) > \enc{\prob_\nn}
      \mright)
    }\right) \\
    \gtrsim_\nn
    \prob_\nn \cdot
    \EE_\nn\left(\quot{
      \enc{\ctsind{\delta_\nn}}\mleft(
        \enc{\pt}_{\enc{\deff}(\enc{\nn})}(\enc{\phi_\nn}) > \enc{\prob_\nn}
      \mright)
    }\right).
\end{multline*}
  \proofin{\ref{app:st}}
\end{restatable}
\noindent The indicator variable $\OneOperator(\phi)$ represents 1 if $\phi$ is true and 0 if $\phi$ is false.  The continuous indicator variable $\ctsind{\delta}(X>p)$ is an ordinary indicator of the event $X>p$, except that instead of a discontinuity at $X=p$, the value is linear in $X$ on a region of  length $\delta$. Thus the self-trust property gives $\pt_n$ only continuous (limited precision) access to the beliefs of $\pt_{f(n)}$; except for this subtlety, we could have written the more recognizable (but false and undesirable!) statement
  \[
    \pt_\nn\left(\quot{
      \enc{\phi_\nn} \wedge
\mleft(
        \enc{\pt}_{\enc{\deff}(\enc{\nn})}(\enc{\phi_\nn}) > \enc{\prob_\nn}
      \mright)
    }\right)
    \gtrsim_\nn
    \prob_\nn \cdot
    \pt_\nn\left(\quot{
        \enc{\pt}_{\enc{\deff}(\enc{\nn})}(\enc{\phi_\nn}) > \enc{\prob_\nn}
    }\right),
  \]
where the conditional $\pt_\nn\left(\quot{
      \enc{\phi_\nn} \mid
        \enc{\pt}_{\enc{\deff}(\enc{\nn})}(\enc{\phi_\nn}) > \enc{\prob_\nn}
    }\right)$ 
    has been rearranged to avoid a potential division by 0.

\renewcommand{\proofin}[1]{} 

\section{Discussion}\label{sec:discussion}

We have proposed the \emph{\lic} as a criterion on the beliefs of deductively limited reasoners, and we have described how reasoners who satisfy this criterion (\emph{logical inductors}) possess many desirable properties when it comes to developing beliefs about logical statements (including statements about mathematical facts, long-running computations, and the reasoner themself).

That said, there are clear drawbacks to the logical inductor we describe in \cite{Garrabrant:2016:li}: it does not use its resources efficiently; it is not a decision-making algorithm (i.e., it does not ``think about what to think about''); and the properties above hold either asymptotically (with poor convergence bounds) or in the limit. Further, it is unclear whether logical inductors have good beliefs about counterpossibilities, and whether they take advantage of old evidence. These are enticing directions for further research.

\renewcommand{\rparenthetical}[1]{}

The authors are particularly interested in tools that help AI scientists attain novel statistical guarantees in settings where robustness and reliability guarantees are currently difficult to come by. For example, consider the task of designing an AI system that reasons about the behavior of computer programs, or that reasons about its own beliefs and its own effects on the world. While practical algorithms for achieving these feats are sure to make use of heuristics and approximations, we believe scientists will have an easier time designing robust and reliable systems if they have some way to relate those approximations to theoretical algorithms that are known to behave well in principle. Modern models of rational behavior are not up to this task: formal logic is inadequate when it comes to modeling self-reference, and probability theory is inadequate when it comes to modeling logical uncertainty. We see logical induction as a first step towards models of rational behavior that work in settings where agents must reason about themselves, while deductively limited.

\subsection{Acknowledgements}

We acknowledge Abram Demski, Benya Fallenstein, Daniel Filan, Eliezer Yudkowsky, Jan Leike, J\'anos Kram\'ar, Nisan Stiennon, Patrick LaVictoire, Paul Christiano, Sam Eisenstat, Scott Aaronson, and Vadim Kosoy, for valuable comments and discussions. We also acknowledge contributions from attendees of the MIRI summer fellows program, the MIRIxLA group, and the MIRI$\chi$ group.

This research was supported as part of the Future of Life Institute (futureoflife.org) FLI-RFP-AI1 program, grant~\#2015-144576.

\sloppy
\bibliographystyle{eptcs}
\bibliography{bibliographydoi} 

\end{document}